\documentclass[aps,pra,twocolumn,nofootinbib,longbibliography]{revtex4-2}
\usepackage{bm,amsmath,amssymb,graphicx}
\usepackage[colorlinks,linkcolor=magenta,urlcolor=blue]{hyperref}
\PassOptionsToPackage{hyphens}{url}
\newcommand{\Dbra}[1]{\left\langle#1\right|}
\newcommand{\Dket}[1]{\left|#1\right\rangle}
\newcommand{\Dbras}[1]{\langle#1|}
\newcommand{\Dkets}[1]{|#1\rangle}

\newcommand{\ave}[1]{\langle#1\rangle}

\begin{document}

\title{Stochastic bra-ket interpretation of quantum mechanics}

\author{Hans Christian \"Ottinger}
\email[]{hco@mat.ethz.ch}
\homepage[]{www.polyphys.mat.ethz.ch}
\affiliation{ETH Z\"urich, Quantum Center and Department of Materials, HCP F 43.1, CH-8093 Z\"urich, Switzerland}

\date{\today}

\begin{abstract}
The stochastic nature of quantum mechanics is more naturally reflected in a bilinear two-process representation of density matrices rather than in squared wave functions. This proposition comes with a remarkable change of the entanglement mechanism: entanglement effects do not originate from superpositions of wave functions, but result from the bilinear structure of density matrices. Quantum interference appears as a multiplicative phenomenon rather than an additive superposition mechanism. We propose two general requirements such that the bilinear representation of density matrices is given in terms of two uniquely defined, identically distributed, Markovian stochastic jump processes. These general ideas are illustrated for the Einstein-Podolsky-Rosen and double-slit experiments. The expression of the stochastic nature of quantum mechanics in terms of random variables rather than their probability distributions facilitates an ontological viewpoint and leads us to a bra-ket interpretation of quantum mechanics.\\[1.5mm]
\end{abstract}

\maketitle

\section{Introduction}
The stochastic nature of quantum mechanics calls for an appropriate setting for describing randomness. An appealing description of quantum states is provided by a density matrix $\rho$ on a Hilbert space for a system of interest, by which the average of any observable $A$, which is a self-adjoint operator on the Hilbert space, can be obtained as a trace, $\ave{A} = {\rm tr}(\rho A)$. In the Schr\"odinger picture, the evolution of the density matrix $\rho_t$ describing the time-dependent state of a quantum system is given by the von Neumann equation
\begin{equation}\label{vonNeumanneq}
   i \hbar \frac{d \rho_t}{dt} = [H,\rho_t] ,
\end{equation}
where $\hbar$ is the reduced Planck constant, $H$ is the Hamiltonian of the system, and the square brackets denote the commutator. Except for its greater flexibility in choosing initial conditions, which are not restricted to pure states, the von Neumann equation is equivalent to the Schr\"odinger equation.

Quantum master equations for density matrices have the advantage that they are perfectly suited not only for describing reversible dynamics, but also for dissipative quantum systems \cite{BreuerPetru,Weiss}. A class of robust quantum master equations for dissipative systems has been obtained by quantizing the geometric structures behind classical nonequilibrium thermodynamics \cite{hco199,hco221}. The same type of thermodynamic master equations has also been found by means of projection-operator methods for coarse-graining reversible Hamiltonian systems \cite{Grabert82}.

The most fundamental equations of nature are generally believed to be reversible or Hamiltonian, whereas dissipation is considered to be an emergent phenomenon. However, as we presently do not know the most fundamental equations of nature and, even if we happened to know them at some point, we could never be sure because new observations could always be made, it might be more appropriate to build all physical theories by default on dissipative equations. One should actually consider reversible equations as an idealization that needs to be justified whenever it might be appropriate. Therefore, density matrices governed by quantum master equations appear to provide the most natural setting for describing the evolution of quantum systems.

Although $\rho_t$ accounts for the stochastic nature of quantum mechanics, the von Neumann equation (\ref{vonNeumanneq}) is a deterministic equation. We here distinguish between \emph{probabilistic tools}, such as density matrices, wave functions or probability densities, and \emph{stochastic quantities}, such as random variables or stochastic processes.

It is the goal of this paper to motivate and elaborate a formulation of quantum mechanics in terms of stochastic jump processes. The density matrix governed by the von Neumann equation (\ref{vonNeumanneq}) can be retrieved as a suitable average from such stochastic processes. Whereas superposition of states occurs inevitably whenever one evolves the Schr\"odinger equation from an initial state that is not an eigenstate of the Hamiltonian, superposition is not a relevant feature of the proposed new formulation. Rather, a strict superselection rule is a characteristic feature of the stochastic bra-ket formulation of quantum mechanics, in which entanglement effects are implemented without superposition.

In the subsequent section, we motivate stochastic jump processes in the context of the hydrogen atom (Section~\ref{secLQM}). We then develop a unique representation of quantum mechanics in terms of stochastic jump processes (Section~\ref{secTPU}). Two experiments, the standard interpretation of which heavily relies on superposition states, are then discussed in the stochastic bra-ket formulation of quantum mechanics: the Einstein-Podolsky-Rosen Gedankenexperiment (Section~\ref{secEPR}) and the double-slit experiment (Section~\ref{secDSE}). Our final conclusions (Section~\ref{secconcl}) focus on the interpretation of quantum mechanics, ontological considerations, and the elimination of paradoxes by constraints for the applicability of quantum mechanics.

\section{Hydrogen atom: quantum mechanics versus quantum field theory} \label{secLQM}
A revealing textbook example of an application of quantum mechanics is the hydrogen atom, which leads to the famous prediction for the line spectrum of atomic hydrogen. For this problem one solves the Schr\"odinger equation for the wave function $\psi(\bm{r},t)$,
\begin{equation}\label{Schroeeq}
   i \hbar \frac{\partial \psi(\bm{r},t)}{\partial t} = H \psi(\bm{r},t) =
   \left( - \frac{\hbar^2}{2m_{\rm e}} \nabla^2
   - \frac{{\rm e}^2}{4 \pi \epsilon_0 |\bm{r}|} \right) \psi(\bm{r},t) ,
\end{equation}
where $m_{\rm e}$ is the mass of the electron, e is the elementary electric charge and $\epsilon_0$ is the vacuum permittivity. Note that the Hamiltonian for this problem contains the Coulomb potential of classical electrostatics. In that sense, quantum mechanics possesses a semi-classical character, which is usually not pointed out. Particularly questionable is the occurrence of electron-proton interactions at a distance, which is a consequence of assuming static interactions. A proper dynamic description of interactions requires quantum field theory.

In a more fundamental treatment of the hydrogen atom, the Coulomb interaction should result from an exchange of soft photons with momenta of the order of the electron momentum and wave lengths of the order of the atomic size between the electron and the proton or, more precisely, between the electron and the quarks in a proton. We then are in the domain of quantum field theory, allowing for the creation and annihilation of photons or various particle-antiparticle pairs. Quantum mechanics for systems with a fixed number of particles can only be an approximate theory that arises in the low-energy limit of quantum field theory, when particle creation and annihilation are suppressed (with additional subtleties associated with massless particles).\footnote{We here use the term \emph{quantum mechanics} for systems with a fixed, finite number of degrees of freedom, whereas \emph{quantum field theory} deals with a variable number of degrees of freedom resulting from particle creation and annihilation processes, including the limit of infinitely many degrees of freedom naturally associated with fields. In our terminology, \emph{quantum theory} includes both mechanics and field theory.}

Quantum mechanics should emerge from the quantum field theory of elementary particles and their interactions by making suitable approximations. This observation suggests that special relativity, when properly taken into account at the field theoretic starting point, should be respected in the approximations leading to quantum mechanics. Of course, it may turn out that there are more fundamental theories of elementary particles and their interactions from which one could obtain both quantum mechanics and quantum field theory by suitable approximations.

Whereas the Coulomb interaction potential in the Schr\"odinger equation (\ref{Schroeeq}) acts continuously in time, quantum field theory suggests that interactions are discrete collision events that occur randomly in time. Between collisions, such as photon emissions and absorptions, there exists a well-defined list of free elementary particles at any time. This situation suggests the use of Fock spaces \cite{Fock32,Teller,hcoqft} and stochastic jump processes for describing quantum systems. The natural Fock basis vectors describe states with well-defined free particle content, where free particles are in momentum eigenstates (we assume regularization by a momentum lattice to avoid the problems associated with a continuous basis; physically this means that the position space is finite).

If a quantum system has a well-defined content of elementary particles characterized by a number of intrinsic properties \cite{hcox4} at any time, a superposition of such states should never occur. We therefore adopt the strict superselection rule that a quantum system should always be described by a multiple of a natural Fock basis vector. In making approximations for passing from quantum field theory to quantum mechanics one should make sure that this fundamental superselection rule is inherited ny the approximate theory. This remark implies that the task of identifying a distinguished basis for quantum mechanics is an essential part of the approximation procedure.

Different options for choosing a distinguished basis may arise even in quantum field theory. For example, in the Schwinger model for electrodynamics in $1+1$ dimensions \cite{Schwinger62,LowensteinSwieca71,KogutSusskind75}, one can switch from the standard Fock space for photons, electrons and positrons to an alternative Fock space based on photons and bound electron-positron pairs (see Section~3.3.7 of \cite{hcoqft}). This switch is similar to regarding protons as elementary particles, and it actually illustrates the emergence of a new distinguished basis via confinement.

When quantum field theory is developed on the Fock space of momentum eigenstates, one cannot know the position of a free quantum particle at a given time. However, whenever collisions between typically three or four particles occur, we know that all the colliding particles are at the same position. This remark explains the characteristic particle tracks observed in collider experiments when a high-energy collision is followed by many low-energy collisions in a detector ruled by a certain correlation structure of collision events \cite{hco243}.

The strict superselection rule of quantum field theory, which should be inherited by quantum mechanics, restricts the situations to which quantum mechanics can be applied in a deep and meaningful way. Whereas Dirac's marvelous quantization recipe formally suggests that we can quantize any classical Hamiltonian system, preferably in canonical coordinates, by replacing classical Poisson brackets by quantum commutators, the superselection rule excludes systems that are not characterized by a list of elementary particles, at least in an effective sense (in particular, dead and alive cats are excluded). A rigid restriction of quantum theory to elementary particles, which would be the other extreme, is unpractical because we will never be sure that we know the truly elementary particles. For example, one should not hesitate to apply quantum mechanics to protons.

\section{Two-process unraveling} \label{secTPU}
Inspired by the discussion of the hydrogen atom, our goal is to formulate quantum mechanics in terms of stochastic jump processes, where interactions are treated as discrete collision events. Therefore, we need a splitting of the full Hamiltonian into free and interacting contributions, $H = H^{\rm free} + H^{\rm int}$. We further assume that there exists a distinguished basis of orthonormal eigenstates $\Dkets{m}$ of the free Hamiltonian $H^{\rm free}$, which are labeled by the natural number $m$. The corresponding eigenvalues of $H^{\rm free}$ be given by $E_m$. Finally, we assume that the strict superselection rule of quantum field theory is inherited by quantum mechanics, that is, the state of the quantum system at any time $t$ is described by a complex multiple of some base vector $\Dkets{m_t}$.

We here follow the strategy of so-called unravelings. The idea is to introduce stochastic processes in Hilbert space such that a density matrix evolving according to the von Neumann equation (\ref{vonNeumanneq}) can be extracted in terms of suitable averages. This idea of passing from probabilistic tools like density matrices to equivalent stochastic processes was originally motivated by computer simulations for dissipative quantum systems \cite{BreuerPetru}, but it is useful also for reversible systems and conceptual clarification. The passage from probabilistic tools to stochastic objects is particularly relevant to ontology \cite{hco243}: We wish to deal with the stochastic objects themselves, not with their probability distributions.

For dissipative systems, unravelings typically employ continuous evolution to represent reversible dynamics and a combination of stochastic jump processes with continuous correction terms to reproduce dissipative dynamics \cite{BreuerPetru}. In the context of dissipative quantum field theory \cite{hcoqft}, it has been proposed to treat also the interactions of reversible systems by jumps. The usefulness of this idea for simulating purely reversible quantum dynamics has been elaborated in detail in \cite{hco251}.

In order to escape the magical procedure of squaring the wave function, we choose a bilinear representation of the density matrix in terms of two stochastic processes. With unravelings in terms of two processes $\Dket{\phi}_t$ and $\Dket{\psi}_t$ in Hilbert space, one wishes to reproduce the density matrix $\rho_t$ by the following expectation evaluated on the underlying probability space,
\begin{equation}\label{unravel}
   \rho_t = E\Big( \Dket{\phi}_t \Dbra{\psi}_t \Big) ,
\end{equation}
where we have used Dirac's bra-ket notation for state vectors (kets) and their duals (bras).\footnote{As bra and ket vectors are duals, one can easily switch back and forth between these two versions. I use the bra-ket distinction only for indicating their position in the dyadic product of the two vectors in the representation of the density matrix (\ref{unravel}).} The use of the dyadic product in (\ref{unravel}) is motivated by the task to construct a tensor from state vectors in Hilbert space. It can also be considered as a stochastic two-process generalization of the density matrix of rank one that is associated with a solution of the Schr\"odinger equation. The expectation may be thought of as an average over trajectories of the stochastic processes.

The commutator in the von Neumann equation (\ref{vonNeumanneq}) implies that there are two terms where the Hamiltonian is positioned on the left and on the right side of the density matrix. For the density matrix $\rho_t = \Dket{\psi}_t \Dbra{\psi}_t$ associated with a solution $\Dket{\psi}_t$ of the  Schr\"odinger equation, these two separate terms result from the product rule. If finite jumps rather than differential changes occur in infinitesimal time intervals, the product rule is no longer applicable. We then need stochastic independence of the increments of the bra and ket processes to achieve the separate appearance of the Hamiltonian on the left and right sides.

With the representation (\ref{unravel}), the average of a quantum observable $A$ can be obtained as an expectation of stochastic matrix elements,
\begin{equation}\label{Aave}
   \ave{A} = {\rm tr}(\rho_t A) = E\big( \Dbra{\psi}_t A \Dket{\phi}_t \big) .
\end{equation}
Instead of a squared deterministic wave function, the expectation of a bilinear form of stochastic wave functions provides the average of any observable $A$. Their stochastic nature arises from spontaneous quantum jumps occurring at random times. In the stochastic averaging procedure, nontrivial phase effects and entanglement can arise from this bra-ket formulation. 

The strict superselection rule has the useful effect of reducing the enormous number of possibilities for constructing the  stochastic jump processes $\Dket{\phi}_t$ and $\Dket{\psi}_t$. It naturally guides us to a construction of piecewise continuous trajectories with interspersed jumps among basis vectors for the two independent, identically distributed stochastic processes. A unique stochastic jump process is constructed in Appendix~\ref{uniqueunravel} and described in the following.
\\[1mm]
\underline{Free evolution between jumps:} If the system between the times $t'$ and $t$ is represented by a multiple of the base vector $\Dket{m}$, the complex prefactor oscillates in time and leads to an overall phase shift given by $-E_m(t-t')/\hbar$, which is in accordance with the free Schr\"odinger equation.
\\[1mm]
\underline{Random jumps:} If the system is represented by a multiple of the base vector $\Dket{m}$, a positive rate parameter $r_m$ characterizes an exponentially decaying probability density for a jump to occur in time. If a jump occurs at time $t$, a transition from $c_t \, \Dket{m}$ to a new state at the time $t+$ is determined by the following stochastic jump rule:
\begin{equation}\label{jumprule}
   c_t \, \Dket{m} \rightarrow c_t \, f_{lm} \Dket{l} \quad \mbox{with} \quad 
   p_{lm} \propto \big| \Dbra{l} H^{\rm int} \Dket{m} \big| ,
\end{equation}
where the transition probabilities $p_{lm}$ satisfy the normalization condition  $\sum_l p_{lm}=1$. 
\\[1mm]\indent
A unique construction of the rate parameter
\begin{equation}\label{rmexpr}
   r_m = \frac{1}{\hbar S_m} \sum_{l} |\Dbras{l} H^{\rm int} \Dkets{m}| .
\end{equation}
and the complex factors $f_{lm} $ associated with transitions in terms of the matrix elements $\Dbra{l} H^{\rm int} \Dket{m}$ is elaborated in Appendix~\ref{uniqueunravel}. The numerical factor $S_m$ in (\ref{rmexpr}) characterizes the magnitude of the complex factors $f_{lm}$ occurring in the jump rule. A unique construction of the jump process is achieved by the postulate that the average magnitude of $f_{lm}$ is unity, so that also these complex factors associated with jumps essentially introduce phase shifts.

We still need to specify the initial conditions for the stochastic processes $\Dket{\phi}_t$ and $\Dket{\psi}_t$. The proper initial conditions for the Einstein-Podolsky-Rosen and double-slit experiments are discussed in Sections \ref{secEPR} and \ref{secDSE}, which illustrate how entanglement effects arise in the two-process unraveling without superpositions. In both cases, also the initial conditions for the two processes of the unraveling are independent random variables. Methods for constructing stochastic unravelings of equilibrium initial states at given temperatures, with special emphasis on the ground state at zero temperature, can be found in \cite{hco248}. 

In summary, we have constructed two independently evolving stochastic processes $\Dket{\phi}_t$ and $\Dket{\psi}_t$, each satisfying a strict superselection rule. The density matrix obtained as a correlation function of these processes according to (\ref{unravel}) satisfies the von Neumann equation (\ref{vonNeumanneq}). This bra-ket formulation offers an alternative interpretation of quantum mechanics.

Note that the total kinetic energy of the free particles is not conserved in collisions. As the unraveling is consistent with the von Neumann equation, it is the total rather than the kinetic energy that is conserved. In such a setting, it might be possible to discuss the time-energy uncertainty principle in a natural way \cite{Briggs08}.

In the next two sections, we discuss the Einstein-Podolsky-Rosen and double-slit experiments in terms of two-process unravelings in order to illustrate the basic ideas. For these discussions we use basis vectors inspired by the Fock space of particle physics with the corresponding superselection rule.

\section{Einstein-Podolsky-Rosen Gedankenexperiment} \label{secEPR}
The Einstein-Podolsky-Rosen (EPR) Gedankenexperiment \cite{EPR35} was designed to reveal the occurrence of actions-at-a-distance as a sign of the incompleteness of quantum mechanics. We consider the EPR experiment in the version for photons, which is appealing for both theoretical arguments and experimental realization \cite{FreedmanClauser72,Aspectetal82,Aspect02ip,Bertlmann90,Albertx,Rickles}. We here prefer photons over electrons because their circular polarization states are characterized with respect to the unambiguous direction of their motion, whereas the electron spin states are usually characterized with respect to a preferred direction in space \cite{hcox4} (because their direction of motion is not Lorentz invariant and is actually undefined in the rest frame). The arbitrariness of the chosen direction of space would have to be addressed with undesirable changes of bases.

Pairs of photons with different wave lengths ($\lambda_1 = 551.3 \, {\rm nm}$ and $\lambda_2 = 422.7 \, {\rm nm}$) moving with the same circular polarization in opposite directions can be created in the decay of properly excited calcium atoms \cite{FreedmanClauser72,Aspectetal82}. Whereas the actual decay occurs through an intermediate state within some $5\,{\rm ns}$, we here use the idealization of a single collision event in which a calcium excitation is annihilated and two photons are created.  

In the discussion of the EPR experiment, one usually focuses entirely on polarizations states and neglects spatial information. Spatial information is associated with the annihilation of a calcium excitation into two photons and with the detection of the photons by collisions in a photomultiplier. In between, we do not have any spatial information. However, one may assume a certain correlation structure for collision events, for example, implying that the two photons from a single excitation hit two equally distant detectors at equal times (in the laboratory). Filters can be introduced such that photons with wave length $\lambda_1$ arrive only in one detector, say the left one, whereas photons with wave length $\lambda_2$ arrive only in the other detector, say the right one.

\begin{figure}
\centerline{\includegraphics[scale=0.37]{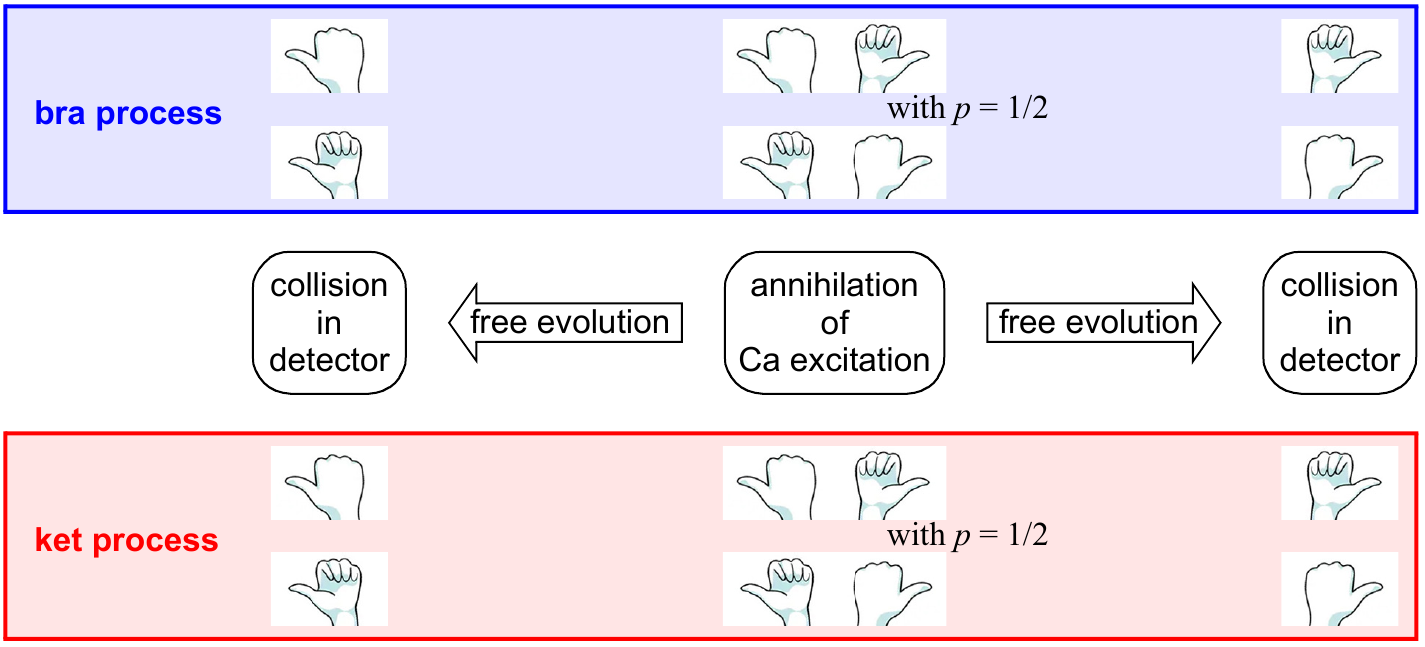}}
\caption[ ]{Schematic illustration of the EPR experiment in the bra-ket interpretation (see text for details).}\label{figEPR}
\end{figure}

The following orthonormal basis vectors for the two-photon states are natural: $\Dket{1} = \Dket{5513,1} \, \Dket{4227,1}$ and $\Dket{2} = \Dket{5513,-1} \, \Dket{4227,-1}$, where each photon is characterized by its wave length in {\AA}ngstr{\"o}m and by its helicity. The eigenvalues for these eigenstates of the free Hamiltonian are independent of helicity, $E_1=E_2$. When a calcium excitation decays, the two processes of the stochastic unraveling are independently initialized with probability $1/2$ in the states $\sqrt{2} \Dket{1}$ or $-\sqrt{2} \Dket{2}$, which are shown as pairs of left or right hands in Figure~\ref{figEPR}. The factors of $\sqrt{2}$ are required because pairs of different vectors in the bra and ket states do not contribute to the normalization of the density matrix or to a signal in the detectors. According to the basic idea of unravelings in (\ref{unravel}), this implies the initial density matrix
\begin{equation}\label{rhoEPR}
   \rho_{\rm EPR} = \frac{1}{2} \, \left( \begin{matrix}
   1 & -1 \\
   -1 & 1
   \end{matrix} \right) ,
\end{equation}
resulting from the independence of the bra and ket states. The pair $(\Dket{\phi}_0, \Dket{\psi}_0)$ randomly takes one of the four combinations $\big(\sqrt{2} \Dket{1}, \sqrt{2} \Dket{1}\big)$, $\big(\sqrt{2} \Dket{1}, -\sqrt{2} \Dket{2}\big)$, $\big(-\sqrt{2} \Dket{2}, \sqrt{2} \Dket{1}\big)$ or $\big(-\sqrt{2} \Dket{2}, -\sqrt{2} \Dket{2}\big)$ with probability $1/4$; on average, this leads to the four matrix elements of (\ref{rhoEPR}). Note that the resulting density matrix is of rank $1$, that is, it corresponds to a pure state in standard quantum mechanics. The pure state associated with the density matrix (\ref{rhoEPR}) is the maximally entangled Bell singlet state $(\Dket{1}-\Dket{2}/\sqrt{2}$, which is the usual starting point for the discussion of the EPR experiment. In the unraveling, this initial condition is produced without any superposition of the natural basis states.

The representation of the density matrix (\ref{rhoEPR}) in the two-process unraveling appears at the center of the schematic illustration of the EPR experiment in Figure~\ref{figEPR}. The photons moving in opposite directions with wave lengths $\lambda_1$ and $\lambda_2$ are indicated by hands with thumbs pointing to the left or right, respectively. The free motion leads to complex phase factors that do not depend on the helicity states, so that the bra-ket average for the density matrix of the helicity states remains unchanged during the free evolution. When the photons hit the detectors at equal distances from the site of their creation, they still have their original, equal helicities: $+1$ and $-1$ with equal probabilities, independently in the bra and ket processes.

Each of the detectors consists of a beam-splitting linear polarizer and two  photomultipliers counting the photons in the two linear polarization states. These polarizers are cubes made of two glass prisms with suitable dielectric thin films on the sides stuck together. The polarizers can be rotated by the angles $\theta_1$, $\theta_2$ around the optical axis with respect to a reference direction. As the photons are moving in opposite directions, also the rotations are performed in opposite directions, so that $\theta_1+\theta_2$ actually is the angle between the two polarizers. 
We are interested in the probabilities for finding the photons in the two optical units in parallel ($\|$) and perpendicular ($\perp$) polarization states. In the natural, intrinsic basis of circular polarization states, these probabilities are the averages of the following observables, which are obtained by transformation from linear to circular polarization states,
\begin{equation}\label{Addef}
   A_{\|\,\|}(\theta_1, \theta_2) = A_{\perp\,\perp}(\theta_1, \theta_2) =
   \frac{1}{4} \, \left( \begin{matrix}
      1 & e^{2 i (\theta_1+\theta_2)} \\
      e^{- 2 i (\theta_1+\theta_2)} & 1 
      \end{matrix} \right)
\end{equation}
and
\begin{eqnarray}
   A_{\|\,\perp}(\theta_1, \theta_2) &=& A_{\perp\,\|}(\theta_1, \theta_2) \nonumber\\ &=&
   \frac{1}{4} \, \left( \begin{matrix}
      1 & - e^{2 i (\theta_1+\theta_2)} \\
      - e^{- 2 i (\theta_1+\theta_2)} & 1
      \end{matrix} \right) . \qquad 
\label{Aodef}
\end{eqnarray}
These averages performed with the density matrix (\ref{rhoEPR}) are used to detect a violation of Bell's inequalities \cite{Bell64,Bell66}, which are nowadays no longer interpreted as actions-at-a-distance but rather as nonlocal correlations in quantum systems \cite{Aspectetal82,Aspect02ip}. Such correlations arise from the multiplicative interplay of the two independent stochastic processes of the bra-ket interpretation. As the density matrices obtained from the maximally entangled Bell singlet state of standard quantum mechanics and from the two-process unraveling coincide, so do the respective predictions for the correlations.

It is important to note that the definition of the observables (\ref{Addef}) and (\ref{Aodef}) implies the following strong \emph{postulate}: When photons pass through macroscopic polarizers, all polarization and phase shift effects coincide with those for classical electromagnetic waves. Also filters are assumed to act on individual photons in the same way as on electromagnetic waves.

\section{Double-slit experiment} \label{secDSE}
\begin{figure}
\centerline{\includegraphics[scale=0.42]{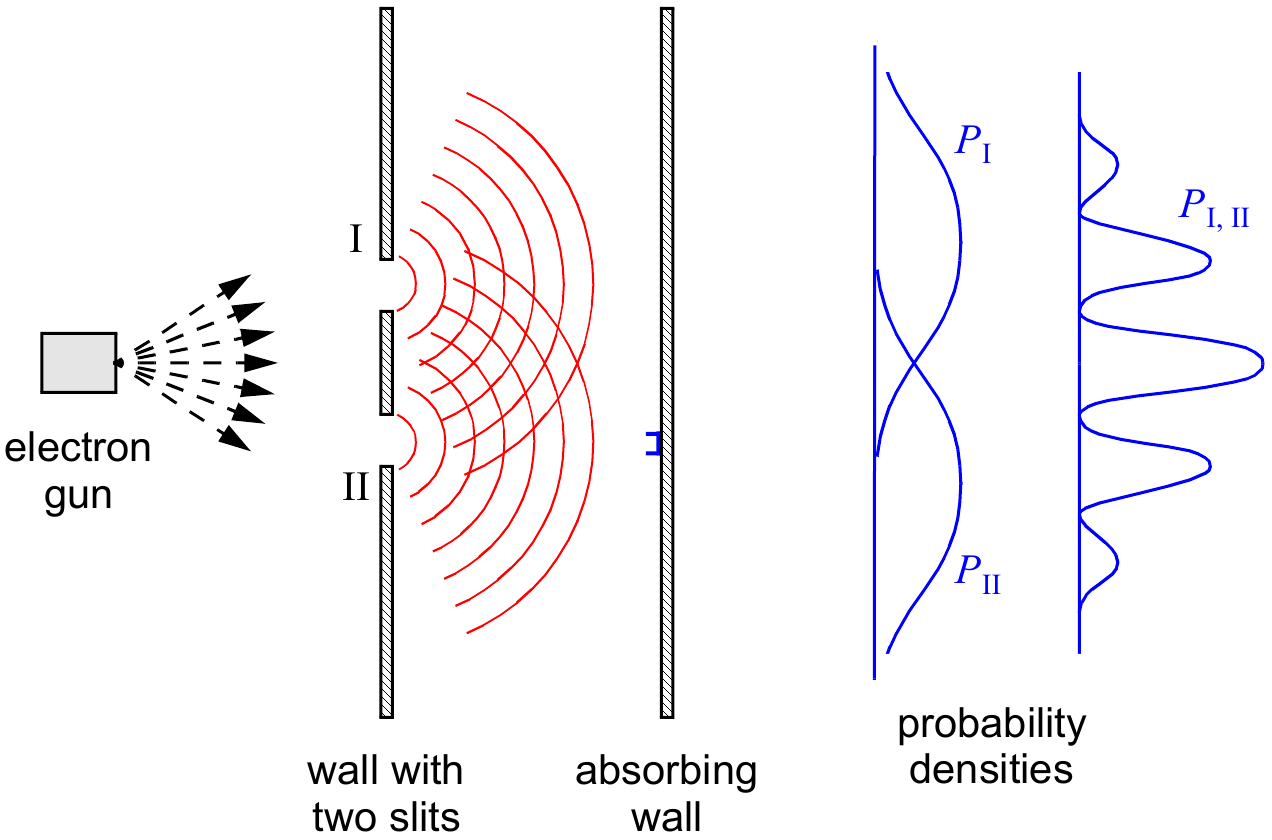}}
\caption[ ]{Double-slit experiment with electrons explained in terms of wave functions (symbolically illustrated by the interference of waves).}\label{figdoubleslit1}
\end{figure}

The double-slit experiment provides most convincing evidence in favor of wave-like properties of quantum particles. As for waves of water or electromagnetic waves, one adds amplitudes and squares the result to obtain intensities exhibiting interference effects. We discuss some details in the spirit of Sections 1-3 to 1-5 in Volume~III of the Feynman Lectures on Physics \cite{Feynman3}. A similar discussion, but from a more philosophical perspective, can be found in Section~7.1 of \cite{Rickles}. An idealized experimental setup for the double-slit experiment with electrons is sketched in Figure~\ref{figdoubleslit1}.

The electron gun in Figure~\ref{figdoubleslit1} could simply be a tungsten wire emitting electrons when heated by an electric current, surrounded by a metal box with a hole in it. A voltage between the wire and the box accelerates the electrons, where all electrons leaving the box through the hole possess nearly the same kinetic energy $E$. The wall with two parallel slits of equal width typically is a thin metal plate. Attached to the absorbing wall, there is a movable electron detector, say a Geiger counter, to measure the distribution of electron hits as a function of position on the absorbing wall.

The electrons are found to hit the absorbing wall as individual events occurring at random positions $x$. If only Slit~I (or II) is open, one finds the probability density $P_{\rm I}$ (or $P_{\rm II}$) for the spatial distribution of events on the absorber, which may also be regarded as intensities (see Figure~\ref{figdoubleslit1}). If both slits are open, one observes the startling interference pattern $P_{\rm I, \, II}$ for the distribution of discrete electron hits on the screen, which is taken as evidence for a ``particle-wave duality.'' Electrons do not simply go through either one or the other slit so that a wave-like pattern with constructive and destructive interference arises. Note that also the single-slit probability densities $P_{\rm I}$ and $P_{\rm II}$ are affected by wave-like diffraction.

\subsection{Standard interpretation}
Following \cite{Feynman3}, the spin of the electron is neglected in our discussion of the double-slit experiment. The probability density $P_{\rm I, \, II}$ obtained with both slits open is clearly not the sum of the probability densities $P_{\rm I}$ and $P_{\rm II}$ for each slit alone. As for waves of water or electromagnetic waves, interference effects can be described by adding amplitudes rather than adding probability densities or intensities,
\begin{equation}\label{amplitudes}
   P_{\rm I, \, II} = | \psi_{\rm I} + \psi_{\rm II} |^2  \quad \neq \quad 
   | \psi_{\rm I} |^2 + | \psi_{\rm II} |^2 = P_{\rm I} + P_{\rm II} ,
\end{equation}
where the complex amplitudes $\psi_{\rm I}$ and $\psi_{\rm II}$ are the wave functions for the single-slit experiments, obtained by solving the corresponding Schr\"odinger equations. The probability densities sketched in Figure~\ref{figdoubleslit1} actually correspond to the intensity profiles obtained from the Fraunhofer diffraction equation of classical wave optics for large distances from the slits. Such a calculation is based on Huygens' idea that every point on a wavefront acts as a source of a spherical wave and that the sum of all these spherical waves determines the further propagation of the wavefront.

Experimental data for cold neutrons can be found in \cite{Zeilingeretal88}. Less detailed experimental data for electrons are available in a much older paper \cite{Jonsson61} (in German), or in its partial translation into English \cite{Jonsson74}.

Explanations of the type sketched above are generally presented and readily accepted as a sound and convincing story of quantum interference by both physicists and philosophers. However, such explanations may be considered as rather symbolic, raising a number of questions. For example, note that $\psi_{\rm I}$ and $\psi_{\rm II}$ in (\ref{amplitudes}) are time-independent wave functions evaluated on the absorbing wall, that is, on a boundary of the domain. On what domain should one actually solve the time-independent Schr\"odinger equation? Between the walls or also around the electron gun? What kind of boundary conditions are required to solve the second-order time-independent Schr\"odinger equation? Is the wave function on the absorbing wall needed as a boundary condition or is it a predictive result of the full solution? Is the sum $\psi_{\rm I} + \psi_{\rm II}$ consistent with the boundary conditions for the double-slit experiment? Furthermore note that the phase shifts, which are at the heart of interference, are in direct correspondence to the flight times for the electrons. How can then a time-independent solution provide the whole story? Why shouldn't small differences in arrival times reveal through which slit an electron has passed? Does a discrete electron hit on the absorbing wall occur at one instant in time or is it smeared in time due to contributions from the two slits? A detailed theoretical discussion that goes far beyond the usual textbook arguments can be found in \cite{Aharonovetal17}.

Maybe the above questions could be addressed most convincingly within Bohmian mechanics \cite{Bohm52,Durretal92,Esfeldetal14,Deckertetal19,Passon}. Whereas all these questions are nonchalantly ignored for the standard interpretation based on wave functions, this is no longer possible for the two-process unraveling.

\subsection{Stochastic bra-ket interpretation}
\begin{figure}
\centerline{\includegraphics[scale=0.42]{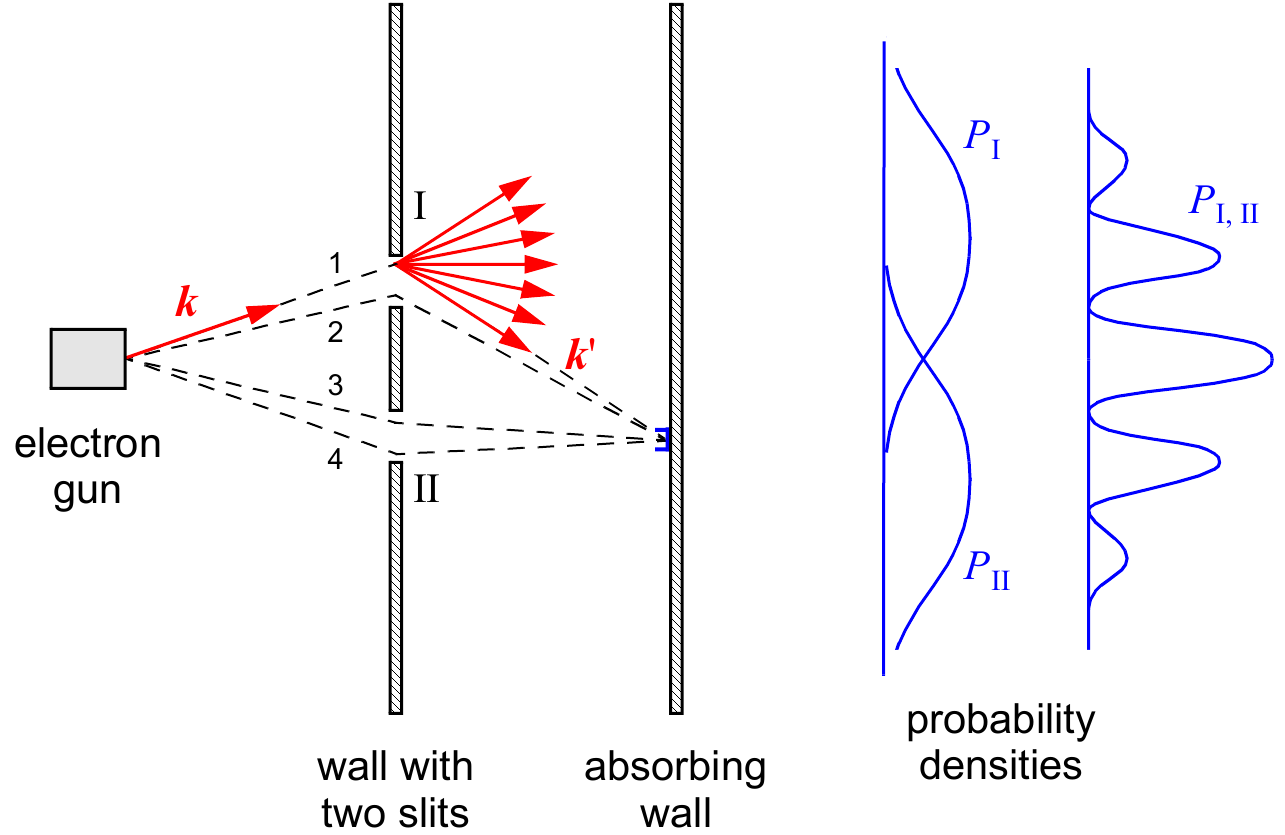}}
\caption[ ]{Double-slit experiment with electrons explained by two-process interpretation.}\label{figdoubleslit2}
\end{figure}

Each of the independent jump processes of the two-process unraveling consists of the following five steps: creation of an electron at the exit hole of the electron gun, free evolution of the electron to one of the two slits, effective collision of the electron in the corresponding narrow slit, free evolution of the electron to the absorbing wall, and absorption of the electron by the wall. As in the standard interpretation, we neglect the electron spin, and we assume that all electrons leave the gun with the same (nonrelativistic) energy $E$. The magnitude of the electron momentum $\bm{k}$ is then restricted by $\bm{k}^2 = 2 m_{\rm e} E$, where $m_{\rm e}$ is the electron mass. We further assume that there is a large but finite number of possible orientations of $\bm{k}$ and that both the set of possible momentum states and their frequencies of occurrence respect the up-down symmetry of the experiment illustrated in Figure~\ref{figdoubleslit2}. Finally, all relevant momenta possess a positive component to the right. 

To elaborate the stochastic jump processes in more detail, we introduce the sets ${\cal K}_{\rm I}, {\cal K}_{\rm II}$ which consist of the momenta pointing from the exit of the electron gun to slit~I or II, respectively, and the set ${\cal K}$ containing all momenta that can occur in an effective elastic collision in a slit. Simultaneously but independently for the two processes of the unraveling, an electron is initiated with a momentum $\bm{k} \in {\cal K}_{\rm I} \cup {\cal K}_{\rm II}$ so that it can pass through one of the slits. After the time determined by the distance of the electron gun from the wall with slits and the component of the momentum $\bm{k}$ normal to the wall, the electron passes a slit. As a result of a high interaction rate with macroscopic matter at the narrow slit, the electron jumps into a state $\bm{k}' \in {\cal K}$, where we make the simplifying assumption of a single effective elastic collision, that is $|\bm{k}'|=|\bm{k}|$. The flight time to the absorbing wall is determined by the distance between the parallel walls and by the normal component of $\bm{k}'$. When the electron hits the absorbing wall, it is assumed to be stopped by an inelastic collision and its momentum jumps to $\bm{0}$.

Equipped with the construction of the independent jump processes of the stochastic unraveling, how can we find the probability for an electron to hit the detector? As the jump process involves only momentum states, we need to have a closer look at the geometry of the experimental setup. For a given position of the detector, we can construct the set ${\cal D}$ of all pairs of momenta $(\bm{k}, \bm{k}')$ for which the electron ends up in the detector. The number of possible momentum states should be so large that, even for a detector with diameter small compared to the width of the slits, the set ${\cal D}$ contains many pairs. Figure~\ref{figdoubleslit2} shows four such $(\bm{k}, \bm{k}')$ pairs and the corresponding piece-wise linear ``trajectories'' that end in the detector. We put the term ``trajectories'' in quotation marks because the position of an electron is known only when collisions occur, that is, in the electron gun, at a slit, and on the absorbing wall, thus defining a $(\bm{k}, \bm{k}')$ pair; in between, the free evolution of the electrons comes with a certain correlation structure between collision events implying flight times and phase factors expressed by piece-wise linearity, just as in the EPR experiment. Two ``trajectories'' per slit is the minimum number required to account for single-slit diffraction. Any finite number of ``trajectories'' per slit can be treated in the same way and even the limit of infinitely many ``trajectories'' may be considered. If, and only if, the ``trajectories'' of both jump processes are contained in ${\cal D}$, there can be a nonzero contribution to the probability for the electron to end up in the detector. Note that the absorption of the electron for the two processes does not occur at exactly the same time; a non-vanishing contribution to the density matrix arises only after both processes have reached the absorbing wall.

The relevance of $(\bm{k}, \bm{k}')$ pairs shows that the calculation of probability densities depends on the sequences of events occurring in the jump processes. Such an analysis goes beyond the evolution of the density matrix according to the von Neumann equation (\ref{vonNeumanneq}) and can only be performed in the context of the more detailed stochastic unravelings of the density matrix. An appealing framework for analyzing such histories of events is provided in \cite{Froehlich21ip}.

We are now in a position to evaluate the probability for an electron to hit the detector. For the pair $j=(\bm{k}, \bm{k}') \in {\cal D}$, the probability $p_j$ is given by the product of the probabilities for generating and scattering the electron in the corresponding momentum states $\bm{k}$ and $\bm{k}'$, respectively. If the complex factors associated with the collisions are assumed to be $1$, the complex phase factor $\phi_j$ is determined by the length of the corresponding piece-wise linear ``trajectory'' or, equivalently, by the flight time. The probability for a pair of independent ``trajectories'' $j,l$ is given by the product $p_j \, p_l$, and the overall probability for electron detection suggested by the matrix element (\ref{Aave}), where $A$ is the indicator function for the pairs $(\bm{k}, \bm{k}')$ for the bra and ket processes to be in ${\cal D}$, is given by\footnote{We do not need to pay attention to the normalization of the density matrix because the number of electrons hitting the slits is not known.}
\begin{equation}\label{2processprob}
   \sum_{j,l \in {\cal D}} p_j p_l \, \phi_j \phi_l^* =
   \bigg| \sum_{j \in {\cal D}} p_j \phi_j \, \bigg|^2 .
\end{equation}
What looks like the square of a sum actually results from the product of all independent combinations of possibilities. The independent processes of the unraveling imply the same result as the corresponding superposition of properly weighted states [in (\ref{amplitudes}), equal weights are assumed]. Without loss of generality, the phase factors at the hole of the electron gun have been taken as unity [random phase factors would average out in (\ref{2processprob})].

It is remarkable that the numerical calculation performed in \cite{Zeilingeretal88} for comparison to experimental diffraction data for neutrons is performed in a fully analogous way. It is also based on the phase factors resulting from the sum of the lengths of the linear paths from the electron source to a point in one of the slits and from there to the detector, and their final result corresponds to the right-hand side of (\ref{2processprob}), thus confirming the assumed correlation structure between collision events. The left-hand side reveals that we here do not obtain this result by superposition, but from the bilinear representation of the density matrix in terms of stochastically independent bra and ket vectors.

\section{Conclusions}\label{secconcl}
The proposed stochastic bra-ket interpretation of quantum mechanics leads to the same equation for density matrices as standard quantum mechanics, which is the von Neumann equation (\ref{vonNeumanneq}). In standard quantum mechanics, this equation for density matrices is inferred from the Schr\"odinger equation for wave functions. In the stochastic bra-ket interpretation, the von Neumann equation is obtained from two stochastic jump processes in Hilbert space. Although the same equation arises in both approaches, they nevertheless are significantly different. The stochastic bra-ket interpretation introduces a new entanglement mechanism and heavily restricts the quantum systems to which the von Neumann equation can be applied, thus reducing the risk of paradoxes.

We have proposed to use unravelings of density matrices in terms of stochastic jump processes to account for the stochastic nature of quantum mechanics in the most direct way. Continuous free Schr\"odinger evolution is interrupted by stochastic quantum jumps that reproduce the proper interaction effects. The bilinear representation (\ref{unravel}) of density matrices in terms of two independently evolving stochastic processes is the most general, natural and versatile option. The enormous freedom in constructing two-process unravelings is massively restricted by a strong superselection rule, which essentially eliminates superposition states from quantum mechanics. This strict superselection rule, which can be justified by obtaining quantum mechanics as a limit of quantum field theory, requires that the content of fundamental particles in a quantum system is well-defined; superpositions of particle states with different characteristic properties are forbidden. In the proposed formulation of quantum mechanics, entanglements no longer arise from superposition, but rather from the dyadic pairing and averaging of the two stochastic state vectors of our unravelings in the fundamental representation (\ref{unravel}) of the density matrix. A unique unraveling is obtained by the additional assumption that the average magnitude of the complex factors associated with jumps is unity.

Superposition states are at the origin of many paradoxes in quantum mechanics. Therefore, a superposition-free implementation of quantum entanglement offers the possibility to eliminate paradoxes. This is a consequence of restricting the applicability of quantum mechanics by the strong superselection rule that results from the origin of quantum mechanics in the quantum field theory of elementary particle physics.

For example, in Schr\"odinger's cat version of the EPR experiment, the superposition state involves a complex macroscopic system. In the tradition of ``Wigner's friend'' \cite{Wigner61ip}, Frauchiger and Renner \cite{FrauchigerRenner18} assume the applicability of quantum mechanics to even more complex superposition states that include agents who are themselves using quantum theory, and they then reveal inconsistencies in conventional quantum mechanics. However, for these situations the applicability of quantum mechanics cannot be justified by making a connection to quantum field theory and verifying the strong superselection rule.

The stochastic jumps of our unravelings reproduce the interaction of standard quantum mechanics. Being equivalent to the interaction part of the Schr\"odinger equation, they reflect the intrinsic stochastic nature of quantum mechanics, not an additional feature. This situation is fundamentally different from the GRW approach \cite{GhirardiRimWeb86}, in which spontaneous collapses of wave functions are an additional stochastic jump feature on top of the full Schr\"odinger dynamics \cite{Bell87ip}. In the GRW approach, superpositions are suppressed only in the passage from microscopic to macroscopic systems \cite{Tumulka06}.

Unlike in the many-worlds interpretation \cite{Everett57,Wallace} with a branching structure into many separate worlds, a combination or overlay of two worlds, or maybe better of two semi-worlds, determines the behavior of a single full world in the bra-ket interpretation. In the many-worlds view, ``world'' is often replaced by ``universe'' whereas, in the bra-ket interpretation, we deal with two stochastic processes in the Hilbert space of a quantum ``system.'' These two processes are rigorously ruled by classical probability theory, where probability is an ontic feature of quantum theory associated with random quantum jumps occurring at random times. The Markov property of its stochastic processes suggests to consider the bra-ket interpretation as a hidden-variable theory. Note that the nature of this hidden-variable theory is very different from deterministic Bohmian mechanics \cite{Bohm52,Durretal92,Esfeldetal14,Deckertetal19,Passon}.

Although it is beyond the scope of this paper to develop a general theory of measurement, it should be pointed out that a theory of measurable multitime correlation functions has been developed in \cite{BreuerPetru}. For the semilinear quantum master equations of thermodynamic origin, which are linear for scalar multiplication but, in general, nonlinear for addition, this theory of measurable correlations has been generalized in \cite{hcoqft}. One can further ask the question whether density matrices can be measured. This question can, for example, be addressed by quantum-state tomography \cite{Rau10,Rau}. Concerning the measurement problem formulated in a concise way by Maudlin \cite{Maudlin95} as the mutual inconsistency of three claims associated with the standard formulation of quantum mechanics in terms of wave functions, none of these three claims is made in the stochastic bra-ket interpretation.

``Density-matrix realism'' as an alternative to ``wave-function realism'' has been discussed in great detail in \cite{Chen21}. We here go a step further and consider the stochastic unraveling of a density matrix as the fundamental representation of a quantum system. Unravelings provide an opportunity of a new interpretation of quantum mechanics, and this new form of realism may be called ``unravelism.'' Complete knowledge of a quantum state requires two stochastic state vectors. These two ``semi-worlds'' are what there ultimately exists, and they play together to characterize the quantum state of the "full world,'' including entanglements. In the context of dissipative quantum field theory, this opportunity has already been explored in \cite{hcoqft,hco243}. The magic number ``two'' in the two-process unraveling corresponds to the number ``2'' in the expression $|\psi|^2$, which relates the wave function to probability. The role of Born's rule is taken over by the bilinear expression (\ref{unravel}) for the density matrix in terms of two more fundamental stochastic processes.

It has been pointed out by Gantsevich \cite{Gantsevich22} that the use of bra-ket pairs is crucial for understanding the particle-wave duality, thus removing the mysteries from quantum mechanics. Gantsevich also describes how classical particles emerge in the bra-ket language when phase effects cancel.

It might appear desirable to couple the two processes of the stochastic bra-ket interpretation (bringing two views of the world together is a challenging task, not only for drunkards). Such a coupling indeed arises as soon as we pass from reversible to irreversible quantum systems \cite{BreuerPetru,hcoqft}. As we argued in the introduction, dissipative dynamics should be considered as the most natural choice, whereas Hamiltonian dynamics can be justified only in exceptional situations. For thermodynamic equilibrium states, it has been observed that the two processes of the unraveling may not be independent \cite{hco248,hco251}. This should not be surprising because thermodynamic equilibrium states are reached by some kind of dissipative process producing entropy, where the precise form or strength of the process is irrelevant.

A quantum field theory that incorporates dissipative smearing at very short length scales has been elaborated in \cite{hcoqft}. It has been argued that the length scale at which dissipative smearing sets a limit to physical resolvability could be identified with the Planck length. As the Planck length involves Newton's gravitational constant it may be concluded that irreversibility is associated with gravity and, therefore, that also the coupling between the two processes of an unraveling might be a gravitational effect. The foundations for including gravity into dissipative quantum field theory have been laid in \cite{hco231,hco252}.

\begin{acknowledgments}
I am grateful to Andrea Oldofredi for many enlightening discussions on the ontological implications of this work. I would like to thank Michael Esfeld, Carlo Rovelli, Sergei Gantsevich, Simon Friederich, Alexei Bazavov, Romain Chessex and Alexander Weyman for helpful comments on previous versions of this paper. I acknowledge inspiring discussions with J{\"u}rg Fr{\"o}hlich and Amine Rusi El Hassani on the operator-algebraic ETH approach (Events--Trees--Histories) to quantum theory.
\end{acknowledgments}

\appendix

\section{Unique unraveling of the von Neumann equation}\label{uniqueunravel}
For an intuitive derivation of a unique stochastic unraveling of the von Neumann equation we rewrite the expression (\ref{unravel}) for the density matrix in terms of two stochastic processes as an average over a large number $N$ of trajectories,
\begin{equation}\label{unravels}
   \rho_t = \frac{1}{N} \sum_{j=1}^N \Dkets{\phi^{(j)}}_t \Dbras{\psi^{(j)}}_t \;,
\end{equation}
in which the bra and ket vectors are multiples of the basis vectors $\Dkets{m}$ at any time,
\begin{equation}\label{multiplesofbasevectors}
   \Dkets{\phi^{(j)}}_t = c^{(j)}_t  \Dkets{m^{(j)}_t} \,, \quad
   \Dkets{\psi^{(j)}}_t = d^{(j)}_t  \Dkets{n^{(j)}_t} .
\end{equation}
According to the von Neumann equation (\ref{vonNeumanneq}), the change of the density matrix in a short time interval $\Delta t$ caused by interactions is given by
\begin{eqnarray}
   i \hbar \Delta \rho_t &=& \frac{1}{N} \sum_{j=1}^N \sum_l \Big[
   \Dbras{l} H^{\rm int} \Dkets{m^{(j)}_t} \, c^{(j)}_t \Dkets{l} \Dbras{\psi^{(j)}}_t 
   \nonumber \\
   &-& \Dkets{\phi^{(j)}}_t \Dbras{l} \, d^{(j)*}_t \; \Dbras{n^{(j)}_t} H^{\rm int} \Dkets{l}
   \Big] \Delta t .
\label{vonNeumannstep}
\end{eqnarray}

In order to find a stochastic unraveling, we characterize a general class of Markovian stochastic jump processes by the following parameters.\\[1mm]
$r_m$: total transition rate for a jump from base state $\Dkets{m}$ to all base states $\Dkets{l}$, where we allow the possibility of an interaction-mediated jump into the state $\Dkets{m}$ with a modified prefactor;\\[1mm]
$p_{lm}$: probability for a transition from base state $\Dkets{m}$ to a particular base state $\Dkets{l}$ when a jump occurs, $\sum_l p_{lm} = 1$;\\[1mm]
$f_{lm}$: complex factor associated with a jump from $\Dkets{m}$ to $\Dkets{l}$.\vspace{1mm}

As a guiding principle for obtaining a unique stochastic jump process we would like to impose the normalization condition that the complex factors $f_{lm}$ are pure phases, $| f_{lm} | = 1$. Pure phase factors would be desirable because they avoid an exponential decay or increase of the norms of the vectors occurring in the stochastic jump processes. However, it turns out that we can only achieve factors with (geometric) average magnitude one ({\sc AMONE}).

In terms of the above parameters of the independent Markovian jump processes, we can write
\begin{eqnarray}
   \Delta \rho_t &=& \frac{1}{N} \sum_{j=1}^N r_{m^{(j)}_t} \Big[ \sum_l
   p_{l m^{(j)}_t} f_{l m^{(j)}_t} \, c^{(j)}_t \Dkets{l} \Dbras{\psi^{(j)}}_t
   \nonumber \\
   &-&  c^{(j)}_t \Dkets{m^{(j)}_t} \Dbras{\psi^{(j)}}_t \Big] \Delta t \nonumber \\
   &+& \frac{1}{N} \sum_{j=1}^N r_{n^{(j)}_t} \Big[ \sum_l
   \Dkets{\phi^{(j)}}_t \Dbras{l} \, d^{(j)*}_t \, p_{l n^{(j)}_t} f^*_{l n^{(j)}_t} \nonumber \\
   &-& \Dkets{\phi^{(j)}}_t \Dbras{n^{(j)}_t} \, d^{(j)*}_t \Big] \Delta t .
\label{jumpstep}
\end{eqnarray}
By matching the first halves of (\ref{vonNeumannstep}) and (\ref{jumpstep}), we obtain the following conditions for the ket process,
\begin{equation}\label{matchket}
   i \hbar r_m (p_{lm} f_{lm} - \delta_{lm}) = \Dbras{l} H^{\rm int} \Dkets{m} .
\end{equation}
From the second halves of (\ref{vonNeumannstep}) and (\ref{jumpstep}), we obtain the complex conjugate version of (\ref{matchket}). Therefore, the jump processes $\Dkets{\phi}_t$ and $\Dkets{\psi}_t$ must be independent, identically distributed processes. Note that the matching procedure is performed in the distinguished basis.

To achieve simple factors $f_{lm}$, we make the ansatz that the transition probabilities are proportional to the matrix elements,
\begin{equation}\label{pchoice}
   p_{lm} \propto |\Dbras{l} H^{\rm int} \Dkets{m}| .
\end{equation}
Note that these transition probabilities are fully determined by the Hamiltonian.
If we introduce the dimensionless weight $W_m$ and the dimensionless normalization factor $S_m$,
\begin{equation}\label{dimlessweights}
   W_m = \frac{1}{\hbar r_m} |\Dbras{m} H^{\rm int} \Dkets{m}| , \quad 
   S_m = \frac{1}{\hbar r_m} \sum_{l} |\Dbras{l} H^{\rm int} \Dkets{m}| ,
\end{equation}
we find
\begin{equation}\label{pmm}
   p_{mm} = \frac{|\Dbras{m} H^{\rm int} \Dkets{m}|}{\sum_{l} |\Dbras{l} H^{\rm int} \Dkets{m}|}
   = \frac{W_m}{S_m} .
\end{equation}

\begin{figure}
\centerline{\includegraphics[scale=0.55]{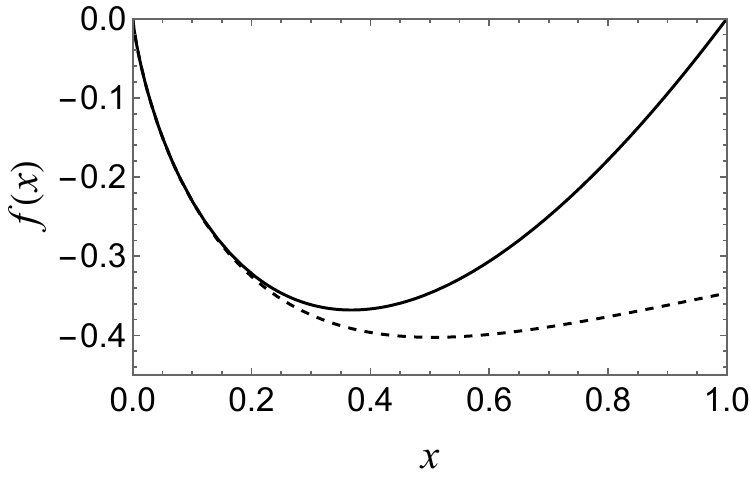}}
\caption[ ]{The functions on the left-hand side (continuous curve) and right-hand side (dashed curve) of the condition (\ref{averagemagnitudex}).}\label{fig_amone}
\end{figure}

For $l \neq m$, the requirement (\ref{matchket}) implies the complex factors
\begin{equation}\label{fchoice}
    f_{lm} = -i \frac{\Dbras{l} H^{\rm int} \Dkets{m}}{| \Dbras{l} H^{\rm int} \Dkets{m} |} \, S_m,
    \qquad | f_{lm} | = S_m .
\end{equation}
Note that the magnitude of these factors is independent of $l$. The further analysis of $f_{mm}$ shows that we cannot simply achieve phase factors by choosing $S_m=1$ because (\ref{matchket}) further implies
\begin{equation}\label{fdchoice}
    f_{mm} = \frac{\hbar r_m - i\Dbras{m} H^{\rm int} \Dkets{m})}{|\Dbras{m} H^{\rm int} \Dkets{m}|}
    \, S_m , \, | f_{mm} | = \frac{\sqrt{1+W_m^2}} {W_m} \, S_m ,
\end{equation}
with a magnitude larger than $S_m$. We therefore impose the {\sc AMONE} condition that the average complex factor should be of magnitude $1$,
\begin{equation}\label{averagemagnitude}
    | f_{mm} |^{p_{mm}} \; S_m^{1-p_{mm}} = 1 .
\end{equation}
This {\sc AMONE} condition can be rewritten in the more explicit form
\begin{equation}\label{averagemagnitudex}
    S_m \ln S_m = W_m \ln \left( \frac{W_m}{\sqrt{1+W_m^2}} \right) .
\end{equation}

\begin{figure}[b]
\centerline{\includegraphics[scale=0.55]{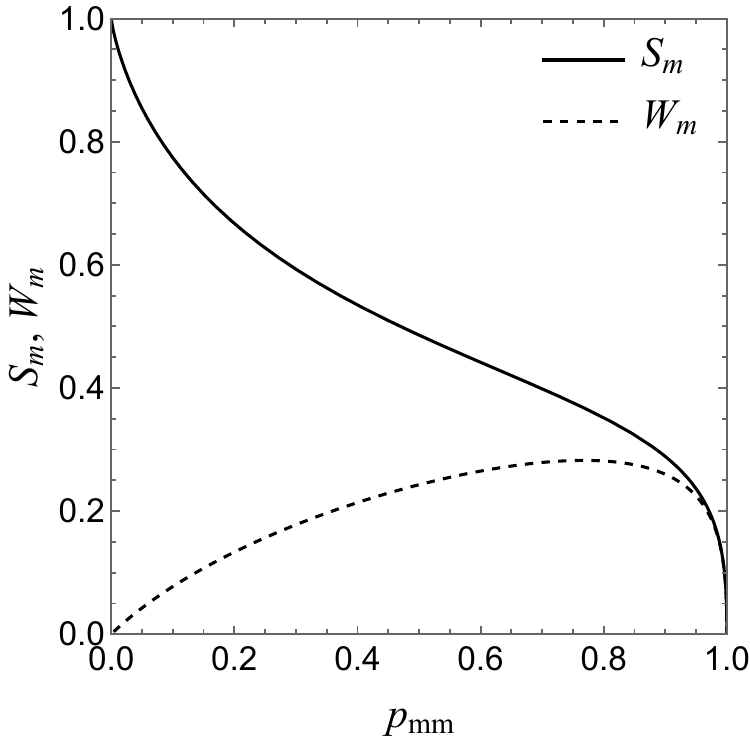}}
\caption[ ]{Explicit solution of the condition (\ref{averagemagnitudex}).}\label{fig_WSsolution}
\end{figure}

The functions on the left- and right-hand sides of the condition (\ref{averagemagnitudex}) are shown in Figure~\ref{fig_amone}. For every value $S_m \in [0,1]$ there is a unique solution $W_m$. We thus obtain the dimensionless normalization factor $S_m$ as the monotonic function of $p_{mm} = W_m/S_m$ shown in Figure~\ref{fig_WSsolution} along with the dimensionless weight factor $W_m$. The parameter $S_m$ plays an important role because it characterizes the magnitudes of the complex factors $f_{lm}$ and, according to (\ref{dimlessweights}), it determines the rate parameter (\ref{rmexpr}).

The limit $p_{mm} \rightarrow 0$ is singular. In the unlikely occurrence of this problem one can add a small constant to the interaction Hamiltonian. The limit $p_{mm} \rightarrow 1$ is irrelevant because no jumps into states different from $\Dkets{m}$ can occur.


%

\end{document}